\documentclass{article}
\usepackage{amsmath}
\usepackage{epsfig}
\usepackage{graphics}
\usepackage{graphicx}
\usepackage{color}
\usepackage{gensymb}

\def\eref{(\ref}
\def\rmd{\mathrm{d}}
\def\rmi{\mathrm{i}}
\def\cal{\mathcal}
\def\rmin{r_{\mathrm{min}}}
\def\rmax{r_{\mathrm{max}}}
\def\Sr{\cal{S}^{\mathrm{r}}}

\def\rv{{\bf r}}
\def\rvp{{\bf r}'}
\def\rvpp{{\bf r}''}

\def\rvh{{\bf \hat r}}
\def\rvph{{\bf \hat r}'}

\def\deps{\rmd \epsilon}

\def\drvh{\rmd {\bf \hat r}}

\def\drvpp{\rmd {\bf r}''}

\def\EF{E_{\mathrm{F}}}
\def\lmax{l_{\mathrm{max}}}

\begin{document}

\title{On the calculation of irregular solutions of the Schr\"odinger equation for non-spherical potentials} 
\author{Rudolf Zeller \\
Institute for Advanced Simulation \\
Forschungszentrum J\"ulich GmbH and JARA \\
D-52425 J\"ulich, Germany}

\date{}

\maketitle

\begin{abstract}
The irregular solutions of the stationary Schr\"odinger equation are
important for the fundamental formal development of scattering theory. They are
also necessary for the analytical properties of the Green function,
which in practice can speed up calculations enormously.
Despite these facts they are seldom considered in numerical
treatments. The reason for this is
their divergent behavior at the origin. This divergence demands high numerical
precision that is difficult to achieve, in particular, for non-spherical potentials
which lead to different divergence rates in the coupled angular momentum channels.
Based on an unconventional treatment of boundary conditions, an
integral-equation method is developed, which is capable to deal with this problem.
The available precision is illustrated by electron-density calculations
for NiTi in its monoclinic B19' structure.
\end{abstract}

\section{Introduction}
\label{sec:int}
The problem of the numerical solution 
of the stationary Schr\"odinger equation
for a single particle has been investigated in
a vast amount of scientific publications, but rarely
with a focus on the irregular solutions. While for spherical
potentials the separation of  radial and angular variables
simplifies the problem into the solution of
one-dimensional radial Schr\"odinger equations,
the situation is more complicated
for non-spherical potentials. Here the separation of
variables leads to radial equations
where different angular momentum components are coupled by the non-diagonal
potential matrix elements. If, as usual, a cutoff is applied
by restricting angular-momentum components to $l \le \lmax$, then $\lmax +1$
independent second-order linear differential equations must be solved
for spherical potentials, while a set of
$(\lmax+1)^2$ coupled second-order linear differential equations
must be solved for non-spherical potentials.
This represents a significant complication, in particular for
the irregular solutions, which diverge with different powers of 
the radial variable $r$ as $r^{-l-1}$.

It is the purpose of this paper to present an approach,
which is capable of treating the divergent behavior in a numerically
efficient manner. The approach is based on the integral-equation method
of Gonzales {\it et al.} \cite{ref:GEK97} who obtained regular solutions of
the radial Schr\"odinger by integrations using 
Clenshaw-Curtis quadratures \cite{ref:CC60}. The numerical solution of
second-order differential equations by integral-equation methods was introduced by
Greengard \cite{ref:G91} and Greengard and Rokhlin \cite{ref:GR91} who pointed
out that stable, high order numerical methods exist for the solution of integral
equations. For instance, evaluation of the integrals with
Clenshaw-Curtis quadratures leads to spectral accuracy.
Spectral accuracy means that the results converge
with the inverse $p$th power of the number of mesh points 
for $p$-times continuously differentiable integrands and
exponentially for $p \rightarrow \infty$.
In contrast to this, the accuracy of solutions of differential equations
by finite difference methods like the Numerov method used in
\cite{ref:HHB10} is limited by a small inverse power of the number of mesh points.

The paper is organized as follows. In Section \ref{sec:math} the mathematical approach
is presented. First the integral equations for the coupled radial equations
are defined and their boundary conditions are discussed.
Then it is explained how the numerical effort can be reduced
by using auxiliary integral equations and how discretization at
Chebyshev collocation points leads to systems
of linear algebraic equations which can be solved by standard numerical
techniques. In Section \ref{sec:numer} two examples are numerically
investigated, a constant potential, for which the results are compared to the analytical results
derived from the expressions given in the appendix,
and a realistic potential as it appears in all-electron density-functional electronic-structure
calculations. It is shown that accurate bound-state wavefunctions and energies are
obtained for constant potentials and that straightforward complex-contour 
integrations for calculating the electron density from the
Green function can be applied. For that purpose, the correct divergence
of the Green function at the origin is enforced 
by using unconventional boundary conditions.
The numerical investigations are done with the
KKRnano code of the JuKKR code package \cite{ref:JuKKR}. This code is based on the
full-potential screened Korringa-Kohn-Rostoker Green-function method \cite{ref:ZDU95}
and was developed for density-functional
calculations for systems with thousands of atoms \cite{ref:TZB12}.

\section{Mathematical approach}
\label{sec:math}

\subsection{Coupled radial equations} 
\label{sec:cou}
The coupled regular and irregular solutions 
of the Schr\"odinger equation 
\begin{equation}
\label{eq:SE}
\left[ - \nabla^2_{\rv} + V (\rv) - E \right] \Psi (\rv;E) = 0
\end{equation}
for the potential $V(\rv)$
can be defined \cite{ref:Z13} by linear Fredholm integral equations 
of the second kind as
\begin{equation}
\label{eq:Rint}
R_{L'L} (r;k) =
j_{l'} (k r) \delta_{L'L} 
+ \int_0^{\infty} \rmd r' r'^2
g_{l'} (r,r';k)
\sum_{L''}
V_{L'L''} (r')
R_{L''L} (r';k)
\end{equation}
and as
\begin{equation}
\label{eq:Sint}
S_{L'L} (r;k) = - \rmi k
h^{(1)}_{l'} (k r) \beta_{L'L} (k)
+ \int_0^{\infty} \rmd r' r'^2
g_{l'} (r,r';k)
\sum_{L''} V_{L'L''} (r') S_{L''L} (r';k) .
\end{equation}
Here
\begin{equation}
\label{eq:VLL}
V_{LL'}(r) = \int \drvh Y_{L} (\rvh) V (\rv ) Y_{L'} (\rvh)
\end{equation}
are matrix elements of the potential and
\begin{equation}
\label{eq:betaLL}
\beta_{L'L} (k) = \delta_{L'L} -
\int_0^{\infty} \rmd r r^2 j_{l'} (k r)
\sum_{L''} V_{L'L''} (r) S_{L''L} (r;k)
\end{equation}
is a matrix,
which implicitly depends on the irregular solutions. The function
$g_{l} (r,r';k)$ is given by
\begin{equation}
\label{eq:freeGF}
g_l (r,r';k) = - \rmi k
\begin{cases}
j_l (k r) h_{l}^{(1)} (k r')
&\text{for}\ r \le r'\\
h_{l}^{(1)} (k r) j_l (k r')
&\text{for}\ r \ge r'. 
\end{cases}
\end{equation}
In these equations and throughout the paper Rydberg atomic units are used.
$j_l$, $h_{l}^{(1)} = j_l + \rmi n_l$ and $n_l$ are spherical Bessel, Hankel 
and Neumann functions,
$Y_L$ spherical
harmonics and $L$ a combined index for the
angular momentum quantum numbers $l$ and $m$. Radial and
angular variables are denoted by $r = |\rv|$ and $\rvh = \rv / r$
and $k = \sqrt E$ is the square root of the energy variable.

The important difference between the inhomogeneous integral equations
\eref{eq:Rint}) for the regular solutions and
\eref{eq:Sint}) for the irregular solutions is that
the source term in \eref{eq:Rint}) contains Bessel functions, which
lead to the $r^{l'}$ behavior of the regular solutions
$R_{L'L} (r;k)$ at the origin, and the
source term in \eref{eq:Sint}) contains Hankel functions,
which lead to the $r^{-l'-1}$ behavior of the irregular solutions
$S_{L'L} (r;k)$ at the origin. The numerical solution of
\eref{eq:Sint}) demands high accuracy because the
integrand contains functions which increase
with different powers $r^{-l''-1} $ at the origin.

If, as it is often done, the coupled radial solutions are not determined from
the Fredholm integral equations \eref{eq:Rint}) and \eref{eq:Sint})
but from differential equations, an additional difficulty arises.
The differential equations can be obtained from
\eref{eq:Rint}) and \eref{eq:Sint})
by applying the operator
\begin{equation}
\label{eq:diffop}
L_r = 
 - \frac{\rmd^2}{\rmd r^2} - \frac{2}{r}
\frac{\rmd}{\rmd r} + \frac{l'(l'+1)}{r^2} - k^2 .
\end{equation}
With $L_r g_{l'} (r',r;k) = - \delta(r-r') / r'^2$,
$L_r j_{l'} (k r) = 0$ and $L_r h^{(1)}_{l'} (k r) = 0$
this leads to the coupled
Schr\"odinger equations
\begin{equation}
\label{eq:Rd}
\sum_{L''} \left[ \left( - \frac{\rmd^2}{\rmd r^2} - \frac{2}{r}
\frac{\rmd}{\rmd r} + \frac{l'(l'+1)}{r^2} - k^2 \right) \delta_{L'L''}
 + V_{L'L''} (r) \right] R_{L''L} (r;k) = 0
\end{equation}
for the regular solutions $R_{L''L} (r;k)$ and to an identical equation
for the irregular solutions $S_{L''L} (r;k)$.
With the cutoff $l \le \lmax$ the
differential equation \eref{eq:Rd}) has $2 (\lmax + 1)^2$ linearly independent 
solutions, one regular and one irregular solution for each value of $L$.
The different solutions are
distinguished by different boundary conditions. These conditions
must be specified explicitly for the differential equation \eref{eq:Rd})
while they are naturally contained in the integral
equations \eref{eq:Rint}) and \eref{eq:Sint}) as a consequence of the source terms.
During the numerical solution of the differential equation \eref{eq:Rd})
it is essential to maintain linear independence of the solutions.
Because of the discretization error,
this represents a considerable challenge already for the regular solutions,
for instance, as it is explained in \cite{ref:EVZ11}, 
and an even greater challenge for the irregular solutions because the
irregular solutions diverge at the origin. 
A discretization error, of course, also occurs in numerical treatments
of integral equations, but by using Clenshaw-Curtis quadrature the error
can be made substantially smaller so that accurate results can be achieved.

\subsection{Boundary conditions}
\label{sec:bou}
For the discussion of the boundary conditions it is convenient to
assume finite integration limits $\rmin$ and $\rmax$. 
This is equivalent
to the assumption that the potential vanishes for $r \le \rmin$ and
$r \ge \rmax$. For such potentials the integral equation
\eref{eq:Sint}) can be written as
\begin{equation}
\label{eq:Sgg3}
\begin{split}
S_{L'L} (r;k) = & - \rmi k h_{l'}^{(1)} (k r) \beta_{L'L} (k) \\
& - \rmi k h_{l'}^{(1)} (k r)
\int_{\rmin}^r \rmd r' r'^2 j_{l'} (k r')
\sum_{L''} V_{L'L''} (r') S_{L''L} (r';k) \\
& - \rmi k j_{l'} (k r)
\int_r^{\rmax} \rmd r' r'^2 h_{l'}^{(1)} (k r')
\sum_{L''} V_{L'L''} (r') S_{L''L} (r';k)
\end{split}
\end{equation}
where \eref{eq:freeGF}) was used. With
\begin{equation}
\label{eq:beta1}
\beta_{L'L} (k) = \delta_{L'L}
- \int_{\rmin}^{\rmax} \rmd r' r'^2 j_{l'} (k r')
\sum_{L''} V_{L'L''} (r') S_{L''L} (r';k) ,
\end{equation}
which arises from \eref{eq:betaLL}) for the finite integration limits,
equation (\ref{eq:Sgg3}) for $S_{L'L} (r;k)$ can be rewritten as
\begin{equation}
\label{eq:Sgg3b}
\begin{split}
S_{L'L} (r;k) = & - \rmi k h^{(1)}_{l'} (k r) \delta_{L'L} \\
& + \rmi k h^{(1)}_{l'} (k r)
\int_r^{\rmax} \rmd r' r'^2 j_{l'} (k r')
\sum_{L''} V_{L'L''} (r') S_{L''L} (r';k)
\\
& - \rmi k j_{l'} (k r)
\int_r^{\rmax} \rmd r' r'^2 h^{(1)}_{l'} (k r')
\sum_{L''} V_{L'L''} (r') S_{L''L} (r';k) .
\end{split}
\end{equation}
This shows that
the irregular solutions can be expressed as
\begin{equation}
\label{eq:Sgg4}
S_{L'L} (r;k) = -\rmi k j_{l'} (k r) C_{L'L} (r;k)
- \rmi k h^{(1)}_{l'} (k r) D_{L'L} (r;k)
\end{equation}
where the matrix functions $C_{L'L} (r;k)$ and $D_{L'L} (r;k)$ are defined as
\begin{equation}
\label{eq:CLpL}
C_{L'L} (r;k) = 
\int_r^{\rmax} \rmd r' r'^2 h^{(1)}_{l'} (k r')
\sum_{L''} V_{L'L''} (r') S_{L''L} (r';k)
\end{equation}
and as
\begin{equation}
\label{eq:DLpL}
D_{L'L} (r;k) = \delta_{L'L} - 
\int_r^{\rmax} \rmd r' r'^2 j_{l'} (k r')
\sum_{L''} V_{L'L''} (r') S_{L''L} (r';k) .
\end{equation}
From \eref{eq:Sgg4}) the inner and outer boundary conditions are obtained as
\begin{equation}
\label{eq:Sinbound}
S_{L'L} (r;k) = - \rmi k j_{l'} (k r) C_{L'L} (\rmin;k) 
- \rmi k h^{(1)}_{l'} (k r) D_{L'L} (\rmin;k)
\end{equation}
for $r \leq \rmin$ and as
\begin{equation}
\label{eq:Soutbound}
S_{L'L} (r;k) =
- \rmi k h^{(1)}_{l'} (k r) \delta_{L'L}
\end{equation}
for $r \geq \rmax$.
Similar to \eref{eq:Sgg4}) the regular solutions can be expressed as
\begin{equation}
\label{eq:Rgg4}
R_{L'L} (r;k) = j_{l'} (k r) A_{L'L} (r;k)
- \rmi k h^{(1)}_{l'} (k r) B_{L'L} (r;k)
\end{equation}
with matrix functions $A_{L'L} (r;k)$ and $B_{L'L} (r;k)$ defined as
\begin{equation}
\label{eq:ALL}
A_{L'L} (r;k) = \delta_{L'L} - \rmi k 
\int_r^{\rmax} \rmd r' r'^2 h^{(1)}_{l'} (k r')
\sum_{L''} V_{L'L''} (r') R_{L''L} (r';k)
\end{equation}
and as
\begin{equation}
\label{eq:BLL}
B_{L'L} (r;k) =
\int_{\rmin}^r \rmd r' r'^2 j_{l'} (k r')
\sum_{L''} V_{L'L''} (r') R_{L''L} (r';k) .
\end{equation}
This can be shown by using \eref{eq:freeGF}) in \eref{eq:Rint}) which results in
\begin{equation}
\label{eq:Rgg3}
\begin{split}
R_{L'L} (r;k) &= 
j_{l'} (k r) \delta_{L'L} \\
&- \rmi k j_{l'} (k r)
\int_{r}^{\rmax} \rmd r' r'^2 h^{(1)}_{l'} (k r')
\sum_{L''}
V_{L'L''} (r')
R_{L''L} (r';k) \\
&- \rmi k h^{(1)}_{l'} (k r)
\int_{\rmin}^{r} \rmd r' r'^2 j_{l'} (k r')
\sum_{L''}
V_{L'L''} (r')
R_{L''L} (r';k) .
\end{split}
\end{equation}
From \eref{eq:Rgg4}) the inner and outer boundary conditions are obtained as
\begin{equation}
\label{eq:Rinbound}
R_{L'L} (r;k) = j_{l'} (k r) A_{L'L} (\rmin;k)
\end{equation}
for $r \leq \rmin$ and as
\begin{equation}
\label{eq:Routbound}
R_{L'L} (r;k) = j_{l'} (k r) \delta_{L'L}
- \rmi k h^{(1)}_{l'} (k r) B_{L'L} (\rmax;k)
\end{equation}
for $r \geq \rmax$.

\subsection{Auxiliary integral equations}
\label{sec:aux}
The use of integral equations instead of
differential equations has been hindered in the past by the much larger
computational work.
When the interval from $\rmin$ to $\rmax$ is discretized by
$N$ mesh points, the integral equations given above
can be converted into systems of linear algebraic equations with the dimension $N$.
Thus the computing time scales as $N^3$ whereas
the computing time to solve linear differential equations 
typically scales only linearly with $N$.
This means that the effort increases with the third power of the
interval length $|\rmax - \rmin|$ for the solution of linear integral equations, but
only linearly with $|\rmax - \rmin|$ for the solution of linear differential equations.

To overcome this problem Greengard and Rokhlin \cite{ref:GR91} pointed out
that the cubic scaling with $|\rmax - \rmin|$ is avoided by dividing the interval
into subintervals and by solving auxiliary integral equations locally in each
subinterval. If the interval $[\rmin,\rmax]$ is divided into
$N$ subintervals $[r_{n-1},r_n]$ with $r_0 = \rmin$ and $r_N = \rmax$
and if $p$ discretization points are used in each subinterval,
the computing time scales as $Np^3$ 
for the solution of the auxiliary integral equations. Thus it 
increases only linearly with the interval length $|\rmax - \rmin|$.
Admittedly, the prefactor $p^3$ can be large, which, however,
is not a serious drawback in view of current computer capabilities.
The method of subintervals is based on the property that the integral equation
\eref{eq:Sgg3b}) for the coupled irregular solutions can be written as
\begin{equation}
\label{eq:Sgg3c}
\begin{split}
S_{L'L} (r;k) = &- \rmi k h^{(1)}_{l'} (k r) D_{L'L} (r_n,k)
- \rmi k j_{l'} (k r) C_{L'L} (r_n,k) \\
&+ \rmi k h^{(1)}_{l'} (k r)
\int_r^{r_n} \rmd r' r'^2 j_{l'} (k r')
\sum_{L''} V_{L'L''} (r') S_{L''L} (r';k)
\\
&- \rmi k j_{l'} (k r)
\int_r^{r_n} \rmd r' r'^2 h^{(1)}_{l'} (k r')
\sum_{L''} V_{L'L''} (r') S_{L''L} (r';k) 
\end{split}
\end{equation}
with
\begin{equation}
\label{eq:CLL}
C_{L'L} (r_n;k) = 
\int_{r_n}^{\rmax} \rmd r' r'^2 h^{(1)}_{l'} (k r')
\sum_{L''} V_{L'L''} (r') S_{L''L} (r';k)
\end{equation}
and 
\begin{equation}
\label{eq:DLL}
D_{L'L} (r_n;k) = \delta_{L'L} - 
\int_{r_n}^{\rmax} \rmd r' r'^2 j_{l'} (k r')
\sum_{L''} V_{L'L''} (r') S_{L''L} (r';k) .
\end{equation}
The idea is to solve \eref{eq:Sgg3c}) separately for each subinterval
with $r$ restricted as $r_{n-1}\leq r \leq r_n$ by introducing auxiliary integral
equations
\begin{equation}
\label{eq:Yn}
\begin{split}
Y^n_{L'L} (r;k) &=
j_{l'} (k r) \delta_{L'L} \\
& - h^{(1)}_{l'} (k r)
\int_r^{r_n} \rmd r' r'^2 j_{l'} (k r')
\sum_{L''} V_{L'L''} (r') Y^n_{L''L} (r';k) \\
& + j_{l'} (k r) 
\int_r^{r_n} \rmd r' r'^2 h^{(1)}_{l'} (k r')
\sum_{L''} V_{L'L''} (r') Y^n_{L''L} (r';k)
\end{split}
\end{equation}
and
\begin{equation}
\label{eq:Zn}
\begin{split}
Z^n_{L'L} (r;k) &=
h^{(1)}_{l'} (k r) \delta_{L'L} \\
&- h^{(1)}_{l'} (k r)
\int_r^{r_n} \rmd r' r'^2 j_{l'} (k r')
\sum_{L''} V_{L'L''} (r') Z^n_{L''L} (r';k) \\
&+ j_{l'} (k r) 
\int_r^{r_n} \rmd r' r'^2 h^{(1)}_{l'} (k r')
\sum_{L''} V_{L'L''} (r') Z^n_{L''L} (r';k)
\end{split}
\end{equation}
The advantage of introducing these local solutions
is that they do not depend 
on the unknown solution $S_{L'L} (r;k)$ in contrast to
\eref{eq:Sgg3c}) which contains  
the matrix functions \eref{eq:CLL}) and \eref{eq:DLL}).
With the local solutions,
which can be obtained numerically
as described in Section \ref{sec:cheby},
the irregular solution can be expressed 
in the interval $[r_{n-1},r_n]$ as
\begin{equation}
\label{eq:s4}
S_{L'L''} (r;k) = - \rmi k 
\sum_{L} \left[ Z^n_{L'L} (r;k) C_{LL''} (r_n;k)
+ Y^n_{L'L} (r;k) D_{LL''} (r_n;k) \right] .
\end{equation}
This can be verified by multiplying
\eref{eq:Yn}) with $- \rmi k D_{LL'''} (r_n;k)$
and \eref{eq:Zn}) with $- \rmi k C_{LL'''} (r_n;k)$,
which yield
\begin{equation}
\label{eq:YnD}
\begin{split}
- \rmi k &Y^n_{L'L} (r;k) D_{LL''} (r_n;k) 
= - \rmi k j_{l'} (k r) D_{L'L''} (r_n;k) \\
&+ \rmi k h^{(1)}_{l'} (k r) 
\int_r^{r_n} \rmd r' r'^2 j_{l'} (k r') \sum_{L''} V_{L'L''} (r')
Y^n_{L''L} (r;k) D_{LL'''} (r_n;k) \\
&- \rmi k j_{l'} (k r)
\int_r^{r_n} \rmd r' r'^2 h^{(1)}_{l'} (k r') \sum_{L''} V_{L'L''} (r')
Y^n_{L''L} (r;k) D_{LL'''} (r_n;k) ,
\end{split}
\end{equation}
\begin{equation}
\label{eq:ZnC}
\begin{split}
- \rmi k &Z^n_{L'L} (r;k) C_{LL''} (r_n;k) 
= - \rmi k h^{(1)}_{l'} (k r) C_{L'L''} (r_n;k) \\
&+ \rmi k h^{(1)}_{l'} (k r) 
\int_r^{r_n} \rmd r' r'^2 j_{l'} (k r') \sum_{L''} V_{L'L''} (r')
Z^n_{L''L} (r;k) C_{LL'''} (r_n;k) \\
&- \rmi k j_{l'} (k r)
\int_r^{r_n} \rmd r' r'^2 h^{(1)}_{l'} (k r') \sum_{L''} V_{L'L''} (r')
Z^n_{L''L} (r;k) C_{LL'''} (r_n;k) .
\end{split}
\end{equation}
Adding \eref{eq:YnD} and \eref{eq:ZnC}, both multiplied with $- \rmi k$ and summed over $L$
leads to an integral equation which compared with \eref{eq:Sgg3c})
contains the same source term and the same kernel. Thus the left and
right sides of \eref{eq:s4}) represent the same function.

It remains to calculate the matrix functions given by \eref{eq:CLL})
and \eref{eq:DLL}). This can be done recursively by recognizing that
these functions satisfy the expressions
\begin{equation}
\label{eq:CLLn}
C_{L'L} (r_{n-1};k) = C_{L'L} (r_n;k) + 
\int_{r_{n-1}}^{r_n} \rmd r' r'^2 h^{(1)}_{l'} (k r')
\sum_{L''} V_{L'L''} (r') S_{L''L} (r';k)
\end{equation}
and 
\begin{equation}
\label{eq:DLLn}
D_{L'L} (r_{n-1};k) = D_{L'L} (r_n;k) -
\int_{r_{n-1}}^{r_n} \rmd r' r'^2 j_{l'} (k r')
\sum_{L''} V_{L'L''} (r') S_{L''L} (r';k)
\end{equation}
and by using \eref{eq:s4}) which shows that the function $S_{L''L} (r';k)$ can be expressed
by the local solutions $Y^n_{L'L} (r;k)$ and $Z^n_{L'L} (r;k)$.
This leads to the integrals
\begin{equation}
\label{eq:qjn}
M^{(hY)}_{L'L} (n;k) = - \rmi k
\int_{r_{n-1}}^{r_n} \rmd r' r'^2 h_{l'}^{(1)} (k r')
\sum_{L''} V_{L'L''} (r')
Y^n_{L''L} (r';k)
\end{equation}
\begin{equation}
\label{eq:qnn}
M^{(hZ)}_{L'L} (n;k) = - \rmi k
\int_{r_{n-1}}^{r_n} \rmd r' r'^2 h_{l'}^{(1)} (k r')
\sum_{L''} V_{L'L''} (r')
Z^n_{L''L} (r';k)
\end{equation}
\begin{equation}
\label{eq:qjj}
M^{(jY)}_{L'L} (n;k) = - \rmi k
\int_{r_{n-1}}^{r_n} \rmd r' r'^2 j_{l'} (k r')
\sum_{L''} V_{L'L''} (r')
Y^n_{L''L} (r';k)
\end{equation}
\begin{equation}
\label{eq:qnj}
M^{(jZ)}_{L'L} (n;k) = - \rmi k
\int_{r_{n-1}}^{r_n} \rmd r' r'^2 j_{l'} (k r')
\sum_{L''} V_{L'L''} (r')
Z^n_{L''L} (r';k)
\end{equation}
which can be evaluated numerically as described 
in Section \ref{sec:cheby}.
By using (\ref{eq:qjn}) to (\ref{eq:qnj}) the expressions
\eref{eq:CLLn}) and \eref{eq:DLLn}) can be written as
recursion relations
\begin{equation}
\label{eq:CLLr}
\begin{split}
C_{L'L} (r_{n-1};k) = C_{L'L} (r_n;k)
&+ \sum_{L''} M^{(hZ)}_{L'L''} (n;k) C_{L''L} (r_n;k) \\
&+ \sum_{L''} M^{(hY)}_{L'L''} (n;k) D_{L''L} (r_n;k)
\end{split}
\end{equation}
\begin{equation}
\label{eq:DLLr}
\begin{split}
D_{L'L} (r_{n-1};k) = D_{L'L} (r_n;k)
&- \sum_{L''} M^{(jZ)}_{L'L''} (n;k) C_{L''L} (r_n;k) \\
&- \sum_{L''} M^{(jY)}_{L'L''} (n;k) D_{L''L} (r_n;k)
\end{split}
\end{equation}
starting from  $C_{L'L} (r_N;k) = 0$ and $D_{L'L} (r_N;k) = \delta_{L'L}$. 

The method of subintervals is particularly advantageous for
potentials with a finite number of discontinuities
in radial direction. If the interval boundaries $r_n$ are adapted to the
discontinuities, the discontinuous behavior is treated without
numerical approximations which means that numerical errors only depend on
the smoothness of the potential within the intervals.
The method of subintervals is also advantageous from a computational
point of view because the auxiliary functions \eref{eq:Yn}) and \eref{eq:Zn})
and then the integrals in (\ref{eq:qjn}) to (\ref{eq:qnj}) can be calculated
efficiently on multi-core processors separately for each value of $n$.

\subsection{Modified boundary conditions}
In order to understand the necessity of modified boundary conditions
for accurate density calculations
it is useful to consider the complex-contour integral
\begin{equation}
\label{eq:densG}
n(\rv) = - \frac{2}{\pi} \lim_{\rvp \rightarrow \rv} {\mathrm{Im}}
\int_{-\infty}^{\EF + \rmi 0^{+}} \deps G(\rv,\rvp;\epsilon).
\end{equation}
which is used to calculate the density from the Green function for the
Schr\"odinger equation \eref{eq:SE}).
Here $\EF$ is the Fermi energy, which determines the total charge, and 
$\rmi 0^{+}$ a positive infinitesimal imaginary quantity, 
which is used to avoid the singularities which exist for real values of $\epsilon$.
Around each atomic position the Green function can be written as
\begin{equation}
\label{eq:GYY}
G(\rv, \rvp ; k ) = \sum_{LL'}
Y_L (\rvh) Y_{L'} (\rvph) 
G_{LL'} (r, r' ; k ) 
\end{equation}
where $k$ is given by $k = \sqrt \epsilon$.
The matrix function $G_{LL'} (r, r' ; k )$ consists of two contributions, a so-called
single-scattering and a so-called multiple-back-scattering part.
While the back-scattering part is determined by coupled regular solutions alone,
the single-scattering part contains 
the divergent irregular solutions in the form \cite{ref:Z13}
\begin{equation}
\label{eq:GSR}
G_{LL'} (r, r' ; k ) = \sum_{L''}
S_{LL''} (r_>;k)
R_{L'L''} (r_<;k) .
\end{equation}
Here $r_<$ and $r_>$ are defined as
$r_<=\min(r,r')$ and $r_>=\max(r,r')$.
The effectiveness of the contour integral \eref{eq:densG}) arises from the fact
that the contour can be chosen in upper half of the complex-$\epsilon$ plane, where
the integrand is an analytical function of $\epsilon$, 
such that with only 20 to 30 mesh points on the contour \cite{ref:ZDD82}
highly accurate density results are obtained even for large
systems with many thousand atoms.

Unfortunately, for the contour integral in \eref{eq:densG}) the irregular solutions are absolutely
necessary to maintain the analytical behavior of the Green function as it is discussed
in the appendix of Ref.~\cite{ref:KA96}.
In a numerical treatment with the standard boundary conditions
\eref{eq:Sinbound}) and \eref{eq:Rinbound}),
the divergent behavior of the matrix function $G_{LL'} (r, r' ; k )$
is given by
\begin{equation}
\label{eq:origin1}
- \rmi k \sum_{L''}
h^{(1)}_{l} (k r) D_{LL''} (\rmin;k) A_{L'L''} (\rmin;k) j_{l'} (k r')
\end{equation}
for $r \leq r' \leq \rmin$. The correct behavior 
\begin{equation}
\label{eq:origin2}
- \rmi k j_l (k r) h^{(1)}_l (k r') \delta_{LL'} 
\end{equation}
is obtained from \eref{eq:Gr1}) derived in the appendix.
Comparison of \eref{eq:origin1}) with \eref{eq:origin2}) shows
that the transpose of the matrix $A (\rmin;k)$ must be equal
to the inverse of the matrix $D (\rmin;k)$. Numerically, this cannot be achieved
because two different integral equations
\eref{eq:Rint}) and \eref{eq:Sint}) are used to calculate these matrices.
A solution for this problem is 
to apply modified irregular and regular solutions defined as
\begin{equation}
\label{eq:Stilde}
\tilde{S}_{L'L}(r;k) = \sum_{L''}
S_{L'L''} (r;k) D^{-1}_{L''L} (\rmin;k)
\end{equation}
\begin{equation}
\label{eq:Rtilde}
\tilde{R}_{L'L}(r;k) = \sum_{L''}
R_{L'L''} (r;k) A^{-1}_{L''L} (\rmin;k)
\end{equation}
These solutions have the inner boundary conditions
\begin{equation}
\label{eq:Stinbound}
\tilde{S}_{L'L} (r;k) = - \rmi k j_{l'} (k r) \sum_{L''} C_{L'L''} (\rmin;k) D^{-1}_{L''L} (\rmin;k)
- \rmi k h^{(1)}_{l'} (k r) \delta_{L'L}
\end{equation}
and
\begin{equation}
\label{eq:Rtinbound}
\tilde{R}_{L'L} (r;k) = j_{l'} (k r) \delta_{L'L}
\end{equation}
for $r \leq \rmin$
such that the divergent part of the matrix function 
\begin{equation}
\label{eq:GtSR}
G_{LL'} (r, r' ; k ) = \sum_{L''}
\tilde{S}_{LL''} (r_>;k)
\tilde{R}_{L'L''} (r_<;k) 
\end{equation}
has the correct behavior given in \eref{eq:origin2}).

The modified irregular solutions can be calculated by
\begin{equation}
\label{eq:st4}
\tilde{S}_{L'L''} (r;k) = - \rmi k 
\sum_{L} \left[ Z^n_{L'L} (r;k) \tilde{C}_{LL''} (r_n;k)
+ Y^n_{L'L} (r;k) \tilde{D}_{LL''} (r_n;k) \right] ,
\end{equation}
which is obtained from \eref{eq:s4}) by multiplication with
the matrix $D^{-1} (\rmin;k)$ from the right. The matrices
$\tilde{C} (r_n;k)$ and $\tilde{D} (r_n;k)$, which are given by
\begin{equation}
\tilde{C}_{LL'} (r_n;k) = \sum_{L''} C_{LL''} (r_n;k) D^{-1}_{L''L'} (\rmin;k)
\end{equation}
and by 
\begin{equation}
\tilde{D}_{LL'} (r_n;k) = \sum_{L''} D_{LL''} (r_n;k) D^{-1}_{L''L'} (\rmin;k) ,
\end{equation}
satisfy the
recursion relations \eref{eq:CLLr}) and \eref{eq:DLLr}) with $C$ and $D$
replaced by $\tilde{C}$ and $\tilde{D}$. The only difference is that the
starting values are changed from 
$C_{L'L} (r_N;k) = 0$ and $D_{L'L} (r_N;k) = \delta_{L'L}$
to $\tilde{C}_{L'L} (r_N;k) = 0$ and 
$\tilde{D}_{L'L} (r_N;k) = D^{-1}_{L'L} (\rmin;k)$. 

The disadvantage of the
recursion for $\tilde{C}$ and $\tilde{D}$ is that the matrix 
$\tilde{D} (r_N;k)$ is known only approximately, for instance, by 
using the numerically obtained result for $D(\rmin;k)$. This minor problem,
however, is offset by the significant advantage that the error is known, which 
arises from the inaccuracy of $D(\rmin;k)$,
from the numerical approximations necessary to solve the auxiliary
integral equations and from roundoff errors. This error is
given by the difference between 
$\tilde{D}^{(1)}_{L'L} (\rmin;k)$, which is the numerical result obtained
after the recursion, and $\delta_{L'L}$, which is the exact result for
$\tilde{D}_{L'L} (\rmin;k)$. The knowledge of 
$\tilde{D}^{(1)}_{L'L} (\rmin;k)$ can be used in a second recursion
starting from a better approximation for $\tilde{D} (r_N;k)$
given as the product of $D^{-1}(\rmin;k)$ and
the inverse of $\tilde{D}^{(1)} (\rmin;k)$. Further improved recursions can be added.
In the present study, where 
electron densities up to $\lmax = 8$ were considered, 
rapid convergence was observed and no more than two or three passes through the
recursion were necessary.

While the straighforward use of repeated recursions 
for $\tilde{C}$ and $\tilde{D}$ successfully deals with numerical
approximations, the problem of roundoff errors requires to modify the
recursion relations such that not $\tilde{D}$ but 
\begin{equation}
\hat{D}_{L'L} (r_n;k) = \tilde{D}_{L'L} (r_n;k) - \delta_{L'L}
\end{equation}
is calculated directly. The modified recursion relations given by 
\begin{equation}
\label{eq:ChLLr}
\begin{split}
\tilde{C}_{L'L} &(r_{n-1};k) = \tilde{C}_{L'L} (r_n;k)
+ M^{(hY)}_{L'L} (n;k) \\
& + \sum_{L''} M^{(hZ)}_{L'L''} (n;k) \tilde{C}_{L''L} (r_n;k)
 + \sum_{L''} M^{(hY)}_{L'L''} (n;k) \hat{D}_{L''L} (r_n;k)
\end{split}
\end{equation}
\begin{equation}
\label{eq:DhLLr}
\begin{split}
\hat{D}_{L'L} &(r_{n-1};k) = \hat{D}_{L'L} (r_n;k)
- M^{(jY)}_{L'L''} (n;k) \\
& - \sum_{L''} M^{(jZ)}_{L'L''} (n;k) \tilde{C}_{L''L} (r_n;k)
  - \sum_{L''} M^{(jY)}_{L'L''} (n;k) \hat{D}_{L''L} (r_n;k)
\end{split}
\end{equation}
lead to a matrix $\hat{D}_{L'L} (\rmin;k)$ with norm 
$||\hat{D}_{L'L} (\rmin;k)|| \ll 1$ such that
needed inverse of $\tilde{D}_{L'L} (\rmin,k)
= \hat{D} (\rmin,k) + \delta_{L'L}$, can be calculated reliably as
$ - \hat{D} (\rmin,k) + \delta_{L'L}$. If necessary,
further reduction of $||\hat{D}_{L'L} (\rmin;k)||$ can be achieved 
by evaluating \eref{eq:ChLLr}) and \eref{eq:DhLLr}) with
extended precision.

\subsection{Numerical treatment}
\label{sec:cheby}
The integral-equation method of
Greengard \cite{ref:G91}, Greengard and Rokhlin \cite{ref:GR91}
and Gonzales {\it et al.} \cite{ref:GEK97}
is based on expansions 
of the potential and the solutions of the local
integral equations \eref{eq:Yn}) and \eref{eq:Zn})
in Chebyshev polynomials $T_m(x) = \cos( m \arccos (x) )$. The method uses 
the property that the integral
\begin{equation}
F(\tau) = \int_\tau^1 \rmd \tau' f(\tau')
\end{equation}
of a function
\begin{equation}
\label{eq:TM}
f(\tau) = \sum_{m=0}^M f_m T_m(\tau)
\end{equation}
can be evaluated at the collocation points
\begin{equation} 
\label{eq:tau_m}
\tau_m = \cos \frac{(2 m + 1)\pi}{2(M+1)} ,
\end{equation} 
by matrix multiplication
\begin{equation}
\label{eq:Ff}
F(\tau_m) = \sum_{m'=0}^M {\cal{T}}_{mm'} f(\tau_{m'}) .
\end{equation}
The collocation points $\tau_m$
are the zeros of the Chebyshev polynomial $T_{M+1}(\tau)$.
The matrix ${\cal{T}}$ is given by the product ${\cal{C}}^{-1} {\Sr} {\cal{C}}$
where ${\cal{C}}$ and $\Sr$ 
are the so-called discrete cosine-transform
and right spectral integration matrices.
The discrete cosine-transform matrix, which is given by
\begin{equation}
\label{eq:Cmm}
{\cal{C}}_{mm'} = T_m(\tau_{m'}) ,
\end{equation}
connects the coefficients $f_m$ of the Chebyshev series \eref{eq:TM})
with values of the function at the zeros \eref{eq:tau_m}) by 
\begin{equation} 
\label{eq:fCf}
f_m = \sum_{m'=0}^M {\cal{C}}_{mm'} f(\tau_{m'}) .
\end{equation}
The right spectral integration matrix as given in \cite{ref:GEK97}
connects the coefficients $F_m$ of the Chebyshev series
\begin{equation}
\label{eq:Tintx}
F(\tau) = \sum_{m=0}^M F_m T_m(\tau)
\end{equation}
with the coefficients $f_m$ by
\begin{equation}
\label{eq:FSf}
F_m = \sum_{m'=0}^M {\Sr}_{\kern -0.35em mm'} f_{m'} .
\end{equation} 
In \eref{eq:Tintx}) the $(M+1)$-th coefficient
is neglected which
is justified because of the fast decay of $F_m$
with increasing $m$ 
for sufficiently smooth functions.
The matrix $\Sr$ has the non-zero elements
${\Sr}_{\!\!00} = 1$, ${\Sr}_{\!\!01} = 1/4$, ${\Sr}_{\!\!10} = -1$, ${\Sr}_{\!\!12} = 1/2$ 
and for $m \ge 2$ the non-zero elements
${\Sr}_{\!\!0m} = 1/(1-m^2)$,
${\Sr}_{\!\!m,m+1} = 1/2m$ and ${\Sr}_{\!\!m,m-1} = -1/2m$.
These values can be obtained by using the integration rules
for Chebyshev polynomials.
By using \eref{eq:Ff}) the local integral equations \eref{eq:Yn})
are approximated by
\begin{equation}
\label{eq:Yle}
Y^n_{L'L}(\tau_m;k) = h^{(1)}_{l'} (k \tau_m) \delta_{L'L} + \frac{r_n - r_{n-1}}{2}
\sum_{m'=0}^M \sum_{L''} {\cal{A}}_{L'L''}^{mm'} Y^n_{L''L} (\tau_{m'};k)
\end{equation}
where the factor $(r_n - r_{n-1})/2$ comes from the substitution
$x = 2(r - r_{n-1})/(r_n - r_{n-1})-1$, which transforms the interval
$[r_{n-1},r_n]$ into $[-1,1]$.
The matrix $\cal{A}$ is given by
\begin{equation}
{\cal{A}}_{L'L''}^{mm'} =
\left[ - h^{(1)}_{l'} (k \tau_m) j_{l'}(k \tau_{m'}) 
+ j_{l'} (k \tau_m) h^{(1)}_{l'}(k \tau_{m'}) \right]
{\cal{T}}_{mm'} \tau^2_{m'} V_{L'L''} (\tau_{m'}) .
\end{equation}
The system \eref{eq:Yle}) of linear equations can be solved efficiently
by standard linear algebra software. It
has dimension $(\lmax+1)^2(M+1)$ and requires a 
computing effort that scales as $(\lmax+1)^6(M+1)^3$.

While, in principle, the subintervals can be chosen arbitrarily,
the choice should be adapted to the divergent behavior 
of the irregular solutions for $r \rightarrow 0$. A suitable choice is 
given by the prescription $r_{n-1} = \alpha r_n$
with $\alpha = (r_0 / r_N)^{1/N} = (\rmin / \rmax)^{1/N}$.
The transformation to the standard expansion interval $[-1,1]$
is obtained by the substitution
$r=\frac{1}{2}r_n \left[ (1-\alpha)\tau + 1 + \alpha \right]$. For inverse
powers of $r$ the substitution leads to
\begin{equation}
\label{eq:chebrl}
\int_{\alpha r_n}^{r_n} \frac{1}{r^l} \rmd r
= \frac{2^{l-1}}{(1-\alpha)^{l-1} r_n^{l-1}} \int_{-1}^1 \frac{1}{(a + \tau)^l} \rmd \tau
\end{equation}
with $a = (1+\alpha)/(1-\alpha)$. Here the integrand and the integration limits for
the integral over $\tau$ do not depend on $n$. Thus,
without changing the Chebyshev-expansion order 
the same relative accuracy is obtained for all intervals.

\section{Numerical investigation}
\label{sec:numer}
The numerical performance is
investigated for two examples, a constant potential, which is analytically
solvable, and a realistic non-spherical potential, which is obtained 
by density-functional electronic-structure calculations
for an ordered nickel-titanium alloy. For the constant
potential it is shown that accurate bound-state energies and wavefunctions
can be obtained from the irregular solutions calculated by the integral-equation
approach and that the error of calculated bound-state
energies decreases exponentially with the order of the Chebyshev expansion.
For the NiTi alloy
it is shown that the irregular solutions obtained can be used 
in complex-contour integrations to calculate the density
from the full-potential multiple-scattering Green function. Thus
such calculations can be done straightforwardly for systems
with many atoms in contrast to other treatments
suggested in the past
\cite{ref:KA96, ref:OA05, ref:RSW11}, which are rather elaborated and unlikely
to be useful for systems with more than a few atoms.

\subsection{Bound states for a constant potential}

The standard method for calculating bound states is based
on the property that regular solutions of the Schr\"odinger
equation vanish at infinity if they are evaluated for the correct bound-state energy.
For trial energies the differential equation is solved from
the inside starting with the correct power $r^l$ at $r=0$
and from the outside starting with zero at a large value of $r$. The bound-state
energy and wavefunction are found if for the chosen trial energy the logarithmic
derivatives of both solutions match continuously at an intermediate value of $r$.
The alternative method suggested here is based of the property
that irregular solutions of the Schr\"odinger equation vanish at the origin
if they are evaluated for the correct bound-state energy.
For a constant potential, because of its spherical symmetry, the irregular
solutions are decoupled $S_{L'L} (r;k) = S_{l'} (r;k) \delta_{L'L}$ and
the inner boundary condition \eref{eq:Sinbound}) contains diagonal matrices 
$C_{L'L} (\rmin;k) = C_{l'} (\rmin;k) \delta_{L'L}$
and 
$D_{L'L} (\rmin;k) = D_{l'} (\rmin;k) \delta_{L'L}$.
The condition for a bound state is then given by $D_{l'} (\rmin;k) = 0$
which eliminates the diverging Hankel functions in \eref{eq:Sinbound}).
In mathematical scattering theory the function
\begin{equation}
\label{eq:Dsph}
D_l (\rmin;k) = 1 -
\int_{\rmin}^{\rmax} \rmd r' r'^2 j_l (k r')
V_0 S_l (r';k)
\end{equation}
is known as Jost function. Its analytical properties in the complex-$k$ plane
are comprehensively discussed in 
Ref.~\cite{ref:N60}, where it is explained that bound states correspond to 
zeros of $D_l (\rmin;k)$ for $k$ values on the positive imaginary axis. The determination
of these zeros is a one-dimensional root-finding problem treated
in the present study by Ridder's method. 

\begin{figure}[h!]
\begin{center}
\includegraphics[width=.49\textwidth,clip]{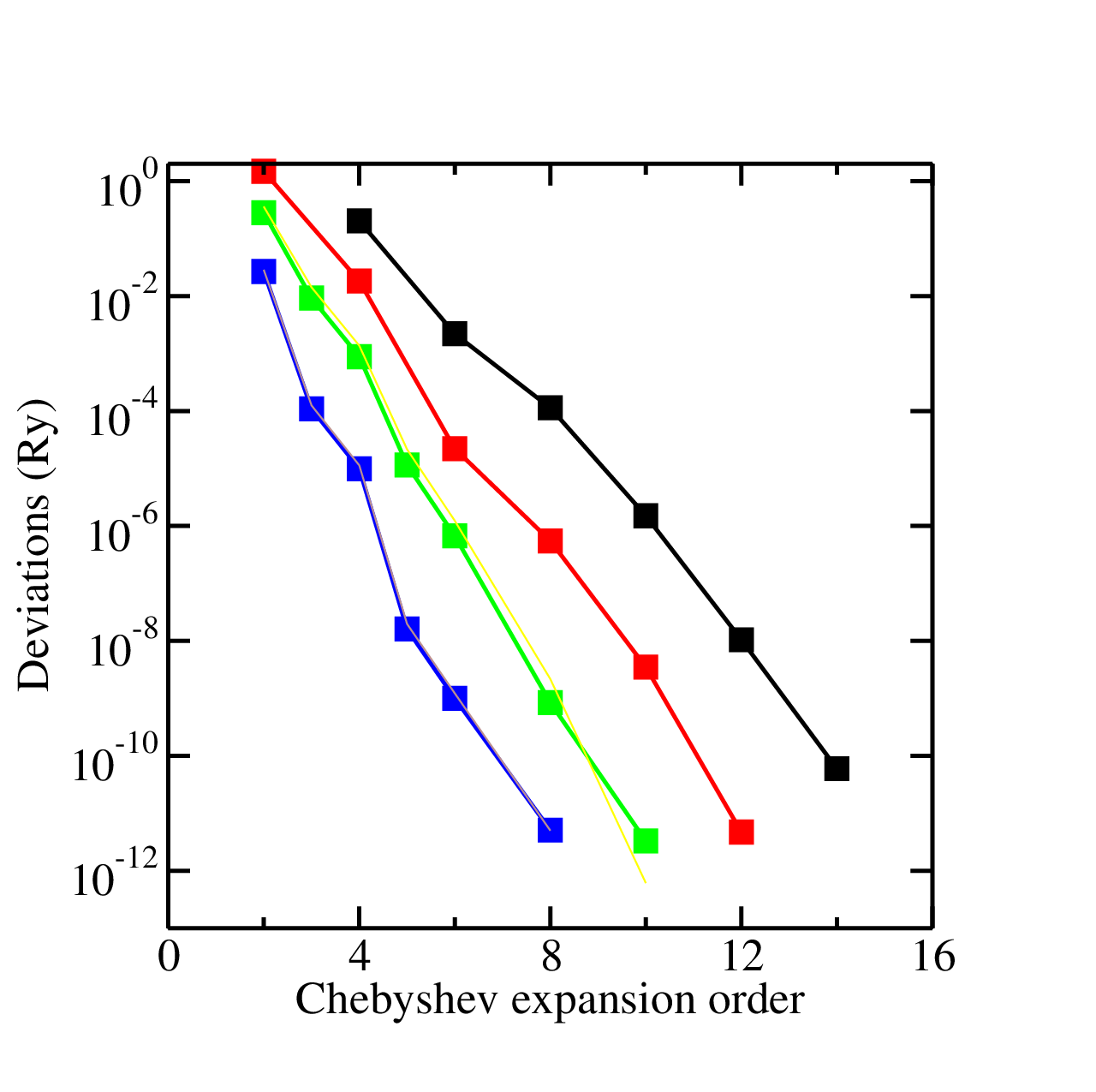}
\raisebox{-1mm}{\includegraphics[width=.49\textwidth,clip]{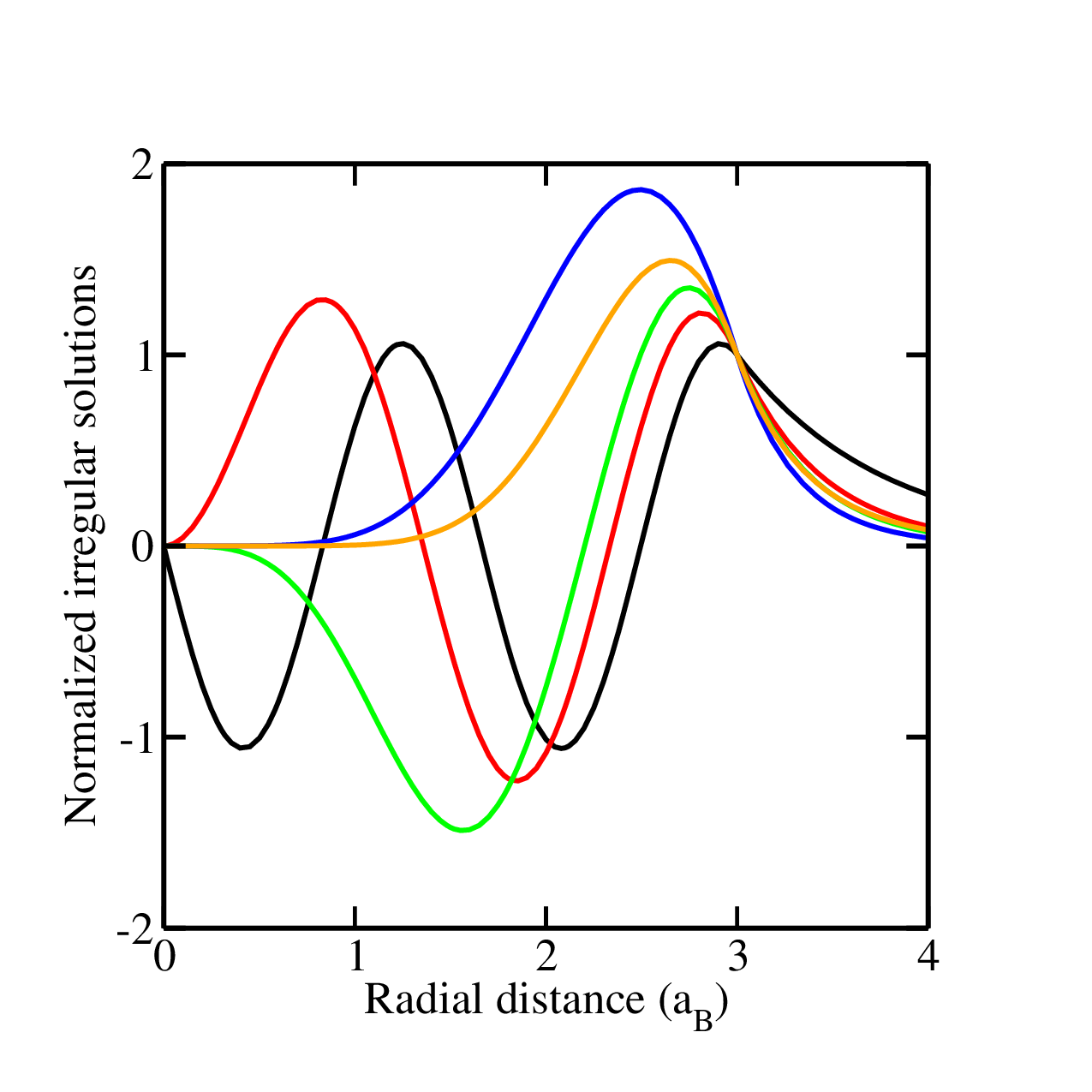}}
\end{center}
\caption{Left graph: Deviations from the exact result for the lowest bound-state energy of $l=0$.
The black, red, green and blue curves are for $N=3$, $N=5$, $N=10$ and $N=30$
intervals.
Right graph: Irregular solutions, multiplied by $r$ and 
normalized to one at $r=3$ ${\mathrm{a}}_{\mathrm{B}}$,
for selected values of $l$. The black, green, red, blue and orange curves 
are for the highest bound-state energies of $l=0$, $l=1$, $l=3$,
$l=5$ and $l=8$.
The constant potential used in the calculations has a depth of -16 Ry and is confined to 
a spherical shell between $r=0.00001$ ${\mathrm{a}}_{\mathrm{B}}$ and
$r=3$ ${\mathrm{a}}_{\mathrm{B}}$.}
\label{fig:wavefunction}
\end{figure}

\begin{table}[h!]
\caption{ Bound-state energies for different $l$ values in Rydberg units for a potential of depth -16 Ry
confined to a spherical shell
between $\rmin = 0.00001$ ${\mathrm{a}}_{\mathrm{B}}$ 
and $\rmax = 3$ ${\mathrm{a}}_{\mathrm{B}}$.}
\begin{tabular}{|r|r|r|r|r|}
\hline
$l$ & $e_1$ & $e_2$ & $e_3$ & $e_4$ \\
\hline
 0 & -15.067032975 & -12.287216857 & -7.738182446 & -1.734147318 \\
\hline
 1 & -14.093970355 & -10.407767360 & -5.037170230 &              \\
\hline
 2 & -12.869064652 & -8.2824156334 & -2.179455290 &              \\
\hline
 3 & -11.405365235 & -5.9291642729 &              &              \\
\hline
 4 & -9.7123791747 & -3.3710840481 &              &              \\
\hline
 5 & -7.7979942918 &               &              &              \\
\hline
 6 & -5.6694973028 &               &              &              \\
\hline
 7 & -3.3343625870 &               &              &              \\
\hline
 8 & -0.8012457212 &               &              &              \\
\hline
\end{tabular}
\label{tab:bs}
\end{table}
Numerical results for bound-state energies and wavefunctions
are shown in table \ref{tab:bs} and figure \ref{fig:wavefunction} for
an attractive potential $V_0 = -16$ Ry, which is confined to a spherical shell
between $\rmin = 0.00001$ ${\mathrm{a}}_{\mathrm{B}}$ and 
$\rmax = 3$ ${\mathrm{a}}_{\mathrm{B}}$. This potential has bound states
up to $l=8$. The energies in table \ref{tab:bs} were obtained using 
$N=10$ intervals of equal length and order $M=10$ for the Chebyshev expansions. 
They deviate by less than $2 \times 10^{-9}$
from the exact energies determined from the zeros of $D_l (\rmin;k)$ using
the analytical expressions given in the appendix. For comparison with values given
in the literature, for example, in Ref.~\cite{ref:AA18} where the same potential
is treated by sinc-interpolants for $\rmin=0$, it is useful
to know that the same digits as in table \ref{tab:bs} are obtained if $\rmin$ is chosen
smaller than $0.00001$ ${\mathrm{a}}_{\mathrm{B}}$.
This is a consequence of the third-order dependence on
$\rmin$ which can be established from the expressions derived in the appendix.

Figure \ref{fig:wavefunction} shows how the energy of the lowest bound
state for $l=0$ converges with the order of the Chebyshev expansion. For the
other bound states the convergence is very similar.
The deviations decrease exponentially with the order and accurate results
are obtained already with a small
number of intervals by using a sufficiently large order. 
A precision of $10^{-10}$ is achieved for $N=3$ intervals
and order $M=14$ which leads to 45 collocation points. 
For a larger number of intervals, where a smaller order gives the same precision, 
more collocation points are necessary. This is, however, not important for the
computational effort,
which scales as $N (M+1)^3$, so that the effects of an increase of $N$ and a decrease $M$
are practically canceled.

Figure \ref{fig:wavefunction} also shows the irregular solutions for the highest
bound-state energies for selected values of $l$
calculated with $N=10$ and $M=8$.
For the presentation they are
multiplied with $r$ and normalized to one at $r=3$ ${\mathrm{a}}_{\mathrm{B}}$.
They are exponentially decaying for $r>3$ ${\mathrm{a}}_{\mathrm{B}}$ and
as a consequence of $D_{l} (\rmin;k) = 0$ regular at $r=0$. Thus they satisfy the requirements
specifying bound-state wavefunctions. It should be noted that different from the
example shown not always the condition $D_{l} (\rmin;k) = 0$ is obtained 
numerically with sufficient precision to conceal the divergent behaviour at
$r=0$. Then it is more appropriate to determine the bound-state wavefunction
not from the irregular solution but
from the regular solution by using the property
that irregular and regular solutions are multiples of one another at the bound-state
energy as discussed in Ref.~\cite{ref:N60}

\subsection{Electron density for NiTi}

Metallic alloys of nickel and titanium have interesting and technologically important
mechanical properties like the shape memory effect. If NiTi in its
low-temperature B19' phase is deformed and heated, it returns to its
original form which persists on cooling. Because of the low symmetry
of the P21/m space group, the spherical-harmonics expansion of the potential
\begin{equation}
V(\rv) = \sum_L V_L(r) Y_L(\rv)
\end{equation}
contains non-zero terms for all values of $l$ in contrast to high symmetry
system like copper or silicon.
The potential for NiTi was determined self-consistently
for $\lmax = 3$.
The exchange-correlation potential was treated in Vosko-Wilks-Nusair parametrization 
\cite{ref:VWN80} and a Monkhorst-Pack grid \cite{ref:MP76} with 16x16x16 points 
was applied for the Brillouin zone integrations. The experimental lattice structure
as given in Ref.~\cite{ref:KTM85} was used with
a = 289.8 pm, b = 464.6 pm, c = 410.8 pm,
{$\gamma = 97.8$\degree} and
Wyckoff (2e) positions ($\pm$0.0372, $\pm$0.6752, 1/4) for Ni and
($\pm$0.4176, $\pm$0.2164, 1/4) for Ti.
The order for the Chebyshev expansion was chosen as $M=8$ and
the subintervals were chosen as follows.
On the outside of the inscribed spheres of the atomic Voronoi cells
the intervals were determined by the kinks of the shape functions
(for an explanation see Ref.~\cite{ref:SAZ90}).
On the inside 
30 intervals were used between $\rmin = 0.00001$ ${\mathrm{a}}_{\mathrm{B}}$ 
and $r = 1.2$ ${\mathrm{a}}_{\mathrm{B}}$ 
with increasing length corresponding to
$\alpha = (0.00001/1.2)^{1/30} = 0.677164$  
and eight intervals
above $r = 1.2$ ${\mathrm{a}}_{\mathrm{B}}$ with
equal length.
The total number of intervals was 64
which leads to overall 576 radial mesh points.
It should be emphasized that the non-spherical potential was used
on all radial mesh points, no cutoff of the non-spherical part near the
atomic centers was applied. Previously, such cutoffs were always necessary
as explained, for instance, in the appendix of Ref.~\cite{ref:Z14a}.

\begin{figure}[h!]
\begin{center}
\includegraphics[width=.49\textwidth,clip]{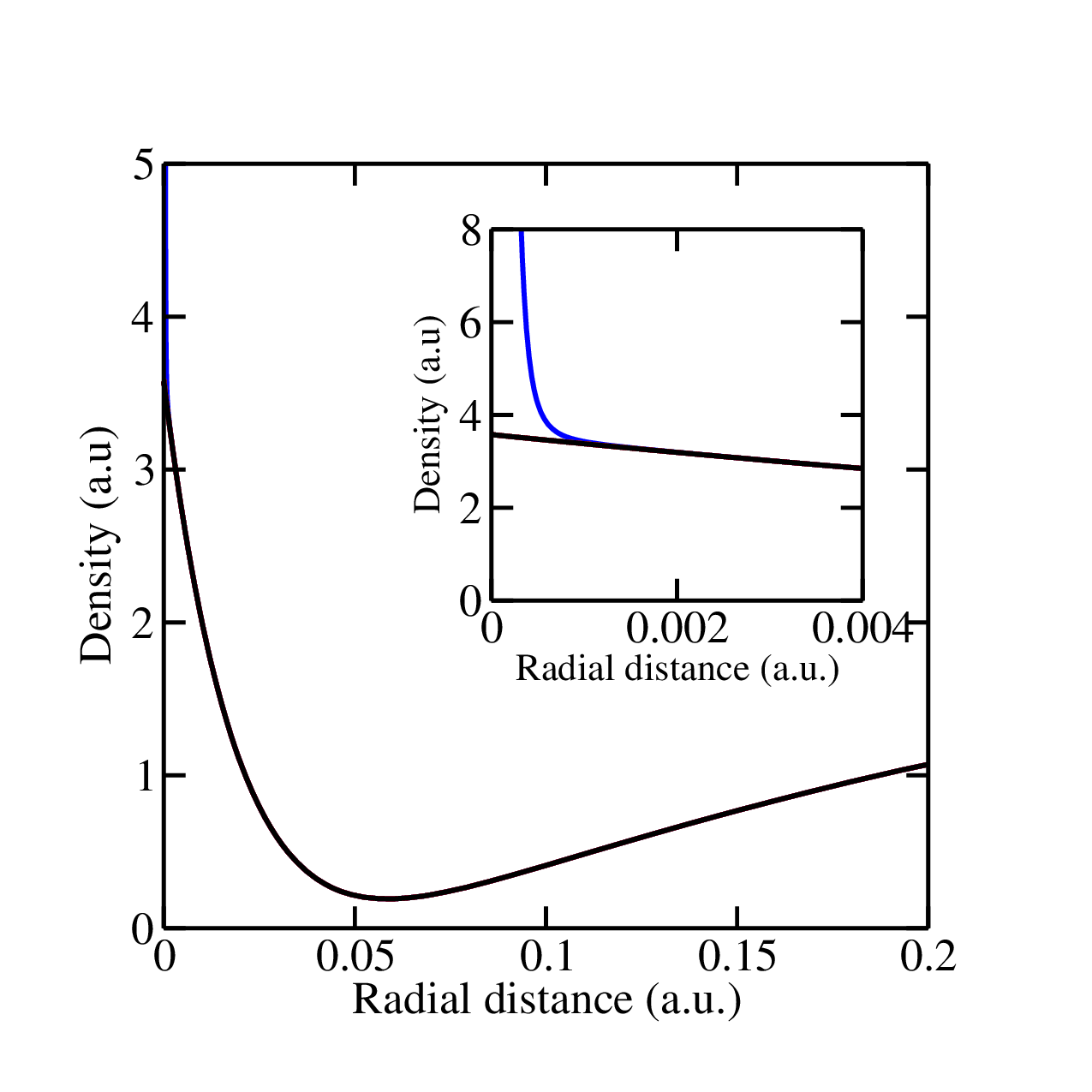}
\includegraphics[width=.49\textwidth,clip]{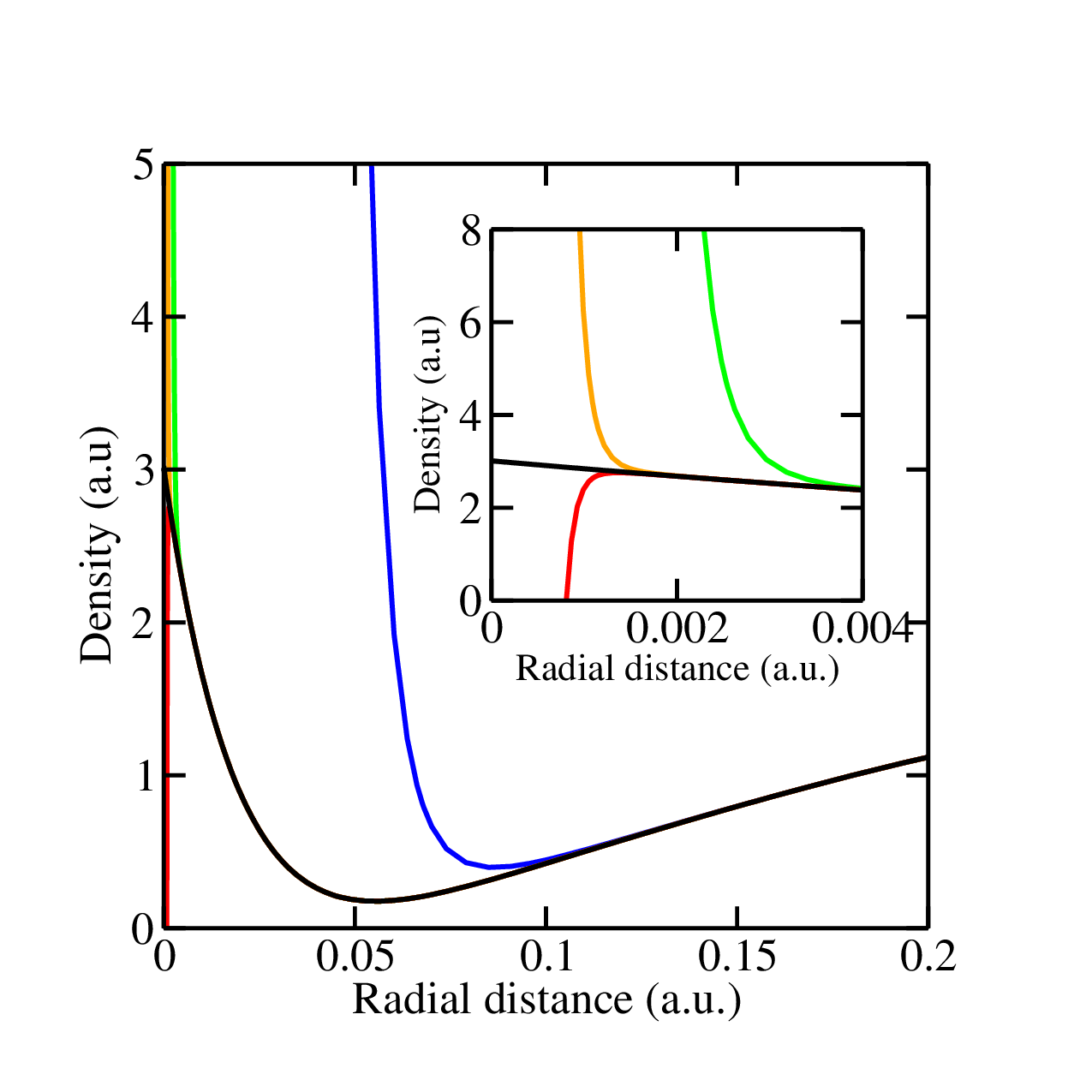}
\end{center}
\caption{ Electron density near the center of a Ni atom in NiTi plotted
in (1,0,0) direction for $\lmax=3$ (left picture) and 
$\lmax=8$ (right picture). The insets display the densities 
on a hundred times smaller range. The different curves are
explained in the text.}
\label{fig:density}
\end{figure}

The self-consistent potential determined in this way was used in calculations
for the electron density for $\lmax \le 8$. In order to save computer resources
a reduced Monkhorst-Pack grid with 6x6x6 points was applied.
Results for the density of the valence electrons near the center of a Ni atom are shown
in figure \ref{fig:density} for $\lmax = 3$ and $\lmax = 8$.
The insets are blowups on a hundred times smaller range. The standard boundary
conditions with the recursion relations \eref{eq:CLLr}) and \eref{eq:DLLr}) lead to
the results shown by blue curves. They deviate from the correct behavior
below $r=0.001$ ${\mathrm{a}}_{\mathrm{B}}$ for $\lmax=3$
and below $r=0.1$ ${\mathrm{a}}_{\mathrm{B}}$ for $\lmax=8$.
These deviations degrade the self-consistency procedure in
density-functional calculations unless somehow they are removed by extrapolation, which
is cumbersome for large systems, or completely neglected
near the atomic centers. Such a neglect might be justified for $\lmax=3$, where the
affected volume is a tiny part of the total volume, but might be unreasonable
for $\lmax=8$, where affected volume is non-negligible.
Straightforward use of the modified boundary conditions leads to
the results show by green curves. In the left picture,
for $\lmax=3$, the green curve, which is hidden under the black curve,
exhibits no divergence at the origin. In the right picture,
for $\lmax=8$, the green curve begins to diverge at 
$r=0.004$ ${\mathrm{a}}_{\mathrm{B}}$, which is considerably smaller than 
$r=0.1$ ${\mathrm{a}}_{\mathrm{B}}$ where the blue curve,
obtained from the unmodified solutions, begins to diverge.
The use of the modified boundary conditions together with one pass through
the modified recursion relations \eref{eq:ChLLr}) and \eref{eq:DhLLr})
leads to the orange curves. In the left picture the orange curve is again 
hidden under
the black curve, while in the right picture it is considerably better than the green curve
by shifting the begin of the divergence from 
$r=0.004$ ${\mathrm{a}}_{\mathrm{B}}$ to
$r=0.0016$ ${\mathrm{a}}_{\mathrm{B}}$. A second pass through
the modified recursion relations \eref{eq:ChLLr}) and \eref{eq:DhLLr})
gives only a small improvement seen in the red curve,
which begins to diverge at about $r=0.0013$ ${\mathrm{a}}_{\mathrm{B}}$.

The question whether better results can be obtained by 
using a larger number of intervals or a higher order of the Chebyshev
expansion was investigated by doubling the number of intervals below
$r = 1.2$ ${\mathrm{a}}_{\mathrm{B}}$ and by increasing the order to
$M=16$. The results obtained in this way are practically identical
to the ones shown in figure \ref{fig:density}. This indicates
that the treatment of the auxiliary integral equations with double precision
floating-point arithmetic is accurate enough. The question whether better
results can be obtained by treating the recursion relations more
accurately was investigated by using quadruple precision instead of double precision.
Then, instead of 
below $r=0.04$ ${\mathrm{a}}_{\mathrm{B}}$,
$r=0.0016$ ${\mathrm{a}}_{\mathrm{B}}$ and 
$r=0.0013$ ${\mathrm{a}}_{\mathrm{B}}$ the density results diverge
only below $r=0.003$ ${\mathrm{a}}_{\mathrm{B}}$,
$r=0.0004$ ${\mathrm{a}}_{\mathrm{B}}$ and 
$r=0.00001$ ${\mathrm{a}}_{\mathrm{B}}$.
This means that divergence-free densities as shown by the black curves can be
obtained for NiTi at least up to $\lmax = 8$.

\section{Summary and outlook}
\label{sec:sum}
An integral-equation approach was presented for the calculation
of irregular solutions of the Schr\"odinger equation for non-spherical
potentials. It was shown how expansions
in Chebyshev polynomials can be used to convert the integral equations
into systems of algebraic equations, which can be solved by standard
software. For that purpose no explicit construction of the Chebyshev series,
but only function values at the zeros of the Chebyshev polynomials are needed.
It was explained that the numerical effort is reduced considerably
by a subinterval technique suggested by Greengard and Rokhlin
and that this technique, with appropriately adapted intervals, is beneficial for potentials
with a finite number of radial discontinuities because
the numerical precision is determined by the smoothness of the
potentials between the discontinuities, but not by the discontinuous behavior.
A numerical investigation was presented for a constant potential
and for a realistic non-spherical potential, which was obtained by density-functional
calculations for a nickel-titanium alloy. It was shown that
accurate bound-state energies can be obtained from the
calculated irregular solutions. It was explained how a precise description
of the divergent behavior of the coupled irregular solutions can be obtained
such that accurate density calculations
by complex-contour integrations are possible.

The approach presented can be extended into several directions.
It is not restricted to the non-relativistic 
Schr\"odinger equation treated here, but useful also for including scalar-relativistic
and spin-orbit-coupling effects \cite{ref:B13} and for full-relativistic calculations
by the Dirac equation \cite{ref:GAK15}. 
It can be extended to calculate bound-state energies for non-spherical potentials,
although with more effort because of near-by roots caused by degeneracy splitting.
It also can be extended to calculate scattering resonances which are determined by zeros
of Jost functions in the complex-$k$ plane. These zeros can be obtained by contour integrals,
for which a Fortran package is available \cite{ref:KVR00}.
Calculations of exchange-correlation and Coulomb potentials,
which require differentiations and integrations, are easily done with
spectral differentiation and integration matrices without introducing
additional numerical approximations. An interesting subject for further research
is the question how much numerical precision is needed to evaluate
the recursion relations for higher
values of $\lmax$ beyond $\lmax = 8$ which was the limit set by the present
capabilities of the applied KKRnano code.

\section{Appendix}

\subsection{Analytical results for constant potentials}
For a potential, which has a constant value $V_0$ for $\rmin \leq r \leq \rmax$
and vanishes for $r < \rmin$ and $r > \rmax$,
the irregular solutions can be calculated analytically. Because of the spherical symmetry
of the potential, the irregular solutions for different
$L$ channels are decoupled and can be written as
\begin{equation}
S_{L'L} (r;k) = S_{l'} (r;k) \delta_{L'L} .
\end{equation} 
With $k_1 = \sqrt {k^2-V_0}$ the functions are given by
\begin{equation}
\label{eq:Sl}
S_l (r;k) = 
\begin{cases}
- \rmi k h^{(1)}_l (k r)
&\text{for}\ r \ge \rmax \\
c_{l} j_{l} (k_1 r) + d_{l} h^{(1)}_{l} (k_1 r)
&\text{for}\ r \le \rmax . 
\end{cases}
\end{equation}
The constants $c_l$ and $d_l$ are determined by the conditions
\begin{equation}
\label{eq:cond1}
c_l j_l (k_1 \rmax) + d_l h^{(1)}_l (k_1 \rmax) = -\rmi k h^{(1)}_l (k \rmax)
\end{equation}
\begin{equation}
\label{eq:cond2}
- k_1 c_l j_{l+1} (k_1 \rmax) - k_1 d_l h^{(1)}_{l+1} (k_1 \rmax) = \rmi k^2 h^{(1)}_{l+1} (k \rmax)
\end{equation}
which guarantee that the solutions are continuous and continuously differentiable at $\rmax$.
Equation (\ref{eq:cond2}) is obtained from \eref{eq:cond1}) by differentiation using the formula
$f'_l(x) = - f_{l+1}(x) + (l/x) f_l(x)$ for derivatives of Bessel and Hankel functions.
The terms arising from $(l/x) f_l(x)$ are omitted in \eref{eq:cond2}) because
they are a simple multiple of \eref{eq:cond1}). Solving \eref{eq:cond1}) and \eref{eq:cond2})
for $c_l$ and $d_l$ leads to
\begin{equation}
\label{eq:cl_V0}
c_l = k k_1 \rmax^2 \left[ k_1 h^{(1)}_{l+1} (k_1 \rmax) h^{(1)}_l (k \rmax) 
- k h^{(1)}_l (k_1 \rmax) h^{(1)}_{l+1} (k \rmax) \right]
\end{equation}
\begin{equation}
\label{eq:dl_V0}
d_l = - k k_1 \rmax^2 \left[ k_1 j_{l+1} (k_1 \rmax)  h^{(1)}_l (k \rmax) 
- k j_l (k_1 \rmax) h^{(1)}_{l+1} (k \rmax) \right]
\end{equation}
Inserting \eref{eq:Sl}) into \eref{eq:Dsph}) 
and using the standard results for integrals of 
products of Bessel and Hankel functions of the same order and 
different arguments leads to
\begin{equation}
D_l(\rmin;k) = D^{(1)}_l + D^{(2)}_l 
\end{equation}
with
\begin{equation}
\label{eq:Dsph1}
\begin{split}
D^{(1)}_l = 1 
& - c_l \left[ k \rmax^2 j_{l+1} (k \rmax) j_l (k_1 \rmax)
- k_1 \rmax^2 j_l (k \rmax) j_{l+1} (k_1 \rmax) \right] \\
& - d_l \left[ k \rmax^2 j_{l+1} (k \rmax) h^{(1)}_l (k_1 \rmax)
- k_1 \rmax^2 j_l (k \rmax) h^{(1)}_{l+1} (k_1 \rmax) \right]
\end{split}
\end{equation}
\begin{equation}
\label{eq:Dsph2}
\begin{split}
D^{(2)}_l = &
  c_l \left[ k \rmin^2 j_{l+1} (k \rmin) j_l (k_1 \rmin)
- k_1 \rmin^2 j_l (k \rmin) j_{l+1} (k_1 \rmin) \right] \\
& + d_l \left[ k \rmin^2 j_{l+1} (k \rmin) h^{(1)}_l (k_1 \rmin)
- k_1 \rmin^2 j_l (k \rmin) h^{(1)}_{l+1} (k_1 \rmin) \right]
\end{split}
\end{equation}
In \eref{eq:Dsph1}) the constants $c_l$ and $d_l$ can be eliminated
by using \eref{eq:cond1}) in the terms
proportional to $k$ and \eref{eq:cond2}) in the terms 
proportional to $k_1$.  
This leads to
\begin{equation}
\label{eq:Dsph1a}
D^{(1)}_l = 1 
+ \rmi k^2 \rmax^2 \left[ j_{l+1} (k \rmax) h^{(1)}_l (k \rmax)
- j_l (k \rmax) h^{(1)}_{l+1} (k \rmax) \right]
\end{equation}
From the Wronskian relation
\begin{equation}
\label{eq:Wrons}
j_{l+1} (x) h^{(1)}_l (x)
- j_l (x) h^{(1)}_{l+1} (x) = \frac{\rmi}{x^2}
\end{equation}
for spherical Bessel functions it follows that $D^{(1)}$ vanishes and
that $D_l(\rmin;k)$ is given by \eref{eq:Dsph2}).

\subsection{Green function at the origin}
The behaviour of the Green function for arguments smaller than $\rmin$
can be determined
from the Dyson equation
\begin{equation}
\label{eq:Dyson}
G(\rv,\rvp;k) = g(\rv,\rvp;k) + \int \drvpp
g(\rv,\rvpp;k) V(\rvpp) G(\rvpp,\rvp;k) .
\end{equation}
by using \eref{eq:GYY}) for
$G(\rv,\rvp;k)$ and the corresponding result
\begin{equation}
\label{eq:gYY}
g(\rv, \rvp ; k ) = \sum_{L}
Y_{L} (\rvh) Y_{L} (\rvph)
g_{l} (r, r' ; k )
\end{equation}
for $g(\rv,\rvp;k)$. This leads to
\begin{equation}
\label{eq:ssgf_Dyson}
\begin{split}
\sum_{LL'}
&Y_L (\rvh) Y_{L'} (\rvph)
G_{LL'} (r, r' ; k )
= \sum_{L}
Y_{L} (\rvh) Y_{L} (\rvph)
g_{l} (r, r' ; k )\\
&+ \sum_{LL'L''}
\int_{\rmin}^{\rmax} \rmd r'' r''^2
Y_{L} (\rvh) g_{l} (r, r'' ; k ) V_{LL''} (r'') 
G_{L''L'} (r'',r';k) Y_{L'} (\rvph) 
\end{split}
\end{equation}
where the integration over the angles was done by using
the definition \eref{eq:VLL}) for the potential matrix elements. 
With the orthogonality of the spherical harmonics the angular coordinates in
\eref{eq:ssgf_Dyson}) can be eliminated, which yields
\begin{equation}
\begin{split}
G_{LL'} (r,r';k) &= 
g_{l} (r,r';k) \delta_{LL'} \\
&+ \int_{\rmin}^{\rmax} \rmd r'' r''^2
g_{l} (r,r'';k)
\sum_{L''} V_{LL''}(r'') G_{L''L'} (r'',r';k) 
\end{split}
\end{equation}
For $r \leq r' \leq \rmin$, the use of
\eref{eq:freeGF}) and \eref{eq:GSR}) leads to
\begin{equation}
\label{eq:Gr0}
\begin{split}
&G_{LL'} (r,r';k) = 
- \rmi k j_l (k r) h^{(1)}_l (k r') \delta_{LL'} \\
& - \rmi k j_l (k r) \sum_{L''L'''} R_{L'L'''} (r';k)
\int_{\rmin}^{\rmax} \rmd r'' r''^2  h^{(1)}_l (k r'')
V_{LL''}(r'') S_{L''L'''} (r'';k) 
\end{split}
\end{equation}
By using \eref{eq:CLpL}) the final result is given by
\begin{equation}
\label{eq:Gr1}
G_{LL'} (r,r';k) 
= - \rmi k j_l (k r) h^{(1)}_l (k r') \delta_{LL'}
- \rmi k j_l (k r) \sum_{L'''} R_{L'L'''} (r';k) C_{LL'''} (\rmin;k)
\end{equation}
Here the only divergent expression is the first term.

\end{document}